\documentclass[prl,aps,twocolumn]{revtex4-1}

\usepackage{amsmath}

\def\beq{\begin{equation}}
\def\eeq{\end{equation}}
\def\bea{\begin{eqnarray}}
\def\eea{\end{eqnarray}}
\def\eq{eq.$\,$\eqref}
\def\Eq{Eq.$\,$\eqref}

\def\det{{\rm det}}
\def\eps{\epsilon}
\def\G{\Gamma}
\def\Gi{\Gamma_\infty}
\def\half{{\textstyle\frac12}}
\def\Im{\mathop{\rm Im}}
\def\Lam{\Lambda}
\def\Nbar{{\overline N}}
\def\Nosc{N_{\rm osc}}

\begin{document}

\title{Nonclassical Degrees of Freedom in the Riemann Hamiltonian}

\author{Mark Srednicki}
\email{mark@physics.ucsb.edu}

\affiliation{Department of Physics, University of California, Santa Barbara, CA 93106}

\begin{abstract}
The Hilbert--P\'olya conjecture states that the imaginary parts of the zeros of the Riemann zeta function are eigenvalues of a quantum hamiltonian.  If so, conjectures by Katz and Sarnak put this hamiltonian in Altland and Zirnbauer's universality class $C$. This implies that the system must have a nonclassical two-valued degree of freedom. In such a system, the dominant primitive periodic orbits contribute to the density of states with a phase factor of $-1$.  This resolves a previously mysterious sign problem with the oscillatory contributions to the density of the Riemann zeros.
 \end{abstract}


\maketitle

The distribution of prime numbers among the integers is a fundamental problem of number theory (see e.g.~\cite{MTB}). It is closely connected to the properties of Dirichlet $L$ functions (including the Riemann zeta function), defined via
\begin{equation}
L(s,\chi) := \sum_{n=1}^\infty \frac{\chi(n)}{n^s}
\label{Lfunc}
\end{equation}
for ${\rm Re}\,s>1$ and by analytic continuation elsewhere, where $\chi(n)$ is a primitive Dirichlet character:
$\chi(n)$ is periodic with smallest period $d$, 
has magnitude one or zero, 
is zero if $d$ and $n$ are not coprime, 
and obeys $\chi(mn)=\chi(m)\chi(n)$.  
The Riemann zeta function is given by $\zeta(s):=L(s,1)$. 
According to the {\it generalized Riemann hypothesis}, any zero of $L(s,\chi)$ with
$0<{\rm Re}\,s<1$ is on the {\it critical line\/} ${\rm Re}\,s=\frac12$; 
these are the {\it nontrivial zeros}, which we will write as $\rho_k = \frac12+i\gamma_k$.  
The generalized Riemann hypothesis implies that each $\gamma_k$ is real, 
and this in turn can be shown to imply that
the number of primes less than $x$ in the arithmetic progression $a, a+d, a+2d, \ldots$ (with $a$ less than
and coprime to $d$) is, in the limit of large $x$,
\begin{equation}
\pi_{a,d}(x)=\frac{1}{\varphi(d)}{\rm Li}(x)+O(x^{1/2+\eps})
\label{Pxnd}
\end{equation}
for all $\eps>0$, where $\varphi(d)$ is the number of integers less than and coprime to $d$ (the Euler totient function), and $\mathop{\rm Li}(x)$ is the logarithmic integral function.  
The exponent of $x$ in the error term increases to 
$\frac12+\mathop{\rm max}\Im\gamma_k+\eps$  
if the generalized Riemann hypothesis is false.   

It is an old idea, now generally known as the {\it Hilbert--P\'olya conjecture} 
(see \cite{D} for a historical review), 
that the nontrivial zeros of each $L$ function are the eigenvalues of an operator (on some Hilbert space) that takes the form $\frac12+iH$, where $H$ is self-adjoint; each $L$ function would have a different $H$.  Furthermore, 
$L(\frac12+iE,\chi)$ is conjectured to be proportional to
the spectral determinant $\det(E-H)$; since the eigenvalues of a hermitian operator must be real, 
the Hilbert--P\'olya conjecture implies the generalized Riemann hypothesis.  
An explicit construction of the $H$'s for the different $L$ functions 
(or, less ambitiously, just for the Riemann zeta function) would therefore be extremely important.  

A large body of analytic and numerical work strongly supports the 
{\it Montgomery--Odlyzko law} (see e.g.~\cite{C}), which states that the statistical distribution of the
$\gamma_k$'s for each $L$ function is the same as the Wigner--Dyson distribution
of the eigenvalues of large hermitian matrices with real diagonal entries and complex off-diagonal entries,
each selected from a gaussian distribution; this is the {\it gaussian unitary ensemble} (GUE) \cite{M}.
A large body of analytic and numerical work also strongly supports the {\it Bohigas--Giannoni--Schmit conjecture} \cite{BGS}, which states that the energy eigenvalues of the hamiltonian for a system that is classically chaotic, and not time-reversal invariant, also obey the GUE distribution.  This leads to the {\it generalized Berry conjecture} \cite{B}: the operator $H$ for each $L$ function
can be obtained by quantizing a classically chaotic system that is not time-reversal invariant.

Katz and Sarnak \cite{KS} have conjectured that $L$ functions corresponding to Dirichlet characters  that are real [$\chi(n)=0,\pm1$] and even [$\chi(-1)=+1$] form a ``family'' 
(that includes the Riemann zeta function) whose members are related (in some fashion) 
by a symplectic symmetry, and furthermore that the spacings of the $\gamma_k$'s for each
member of this family is governed by the distribution of eigenphases of random unitary symplectic matrices.
This agrees with the GUE distribution for $\gamma_k\gg 1$, and predicts a gap in the spectrum near zero; this is well supported by numerical evidence from these $L$ functions \cite{KS,R}. 
Other proposed families have unitary or orthogonal symmetries.

Altland and Zirnbauer \cite{AZ} have classified the possible symmetry classes of quantum
hamiltonians.  The distribution of $\gamma_k$'s found by Katz and Sarnak 
is a predicted property of the energy eigenvalues for a chaotic system in Altland and 
Zirnbauer's class $C$.  We therefore interpret the Katz--Sarnak conjecture, 
in the context of the Hilbert--P\'olya conjecture,
to mean that the quantum system corresponding to the Riemann zeta function
(or any other member of its symplectic family of $L$ functions) should have a hamiltonian
in class $C$.

A hamiltonian in class $C$ takes the form of a generator of $U\!Sp(N)$; more specifically,
\beq
H=A+{\vec\sigma}\!\cdot\!\vec S,
\label{H}
\eeq
where $A$ is a hermitian operator that (when expressed as a matrix in a suitable
basis) is imaginary and antisymmetric, and each $S_i$ ($i=1,2,3$) is a hermitian operator that
(when expressed as a matrix in the same basis) is real and symmetric; finally, 
$\sigma_i$ is a Pauli matrix acting in an additional two-dimensional Hilbert space.  
This extra ``nonclassical two-valuedness" (``klassisch nicht beschreibbare 
Zweideutigkeit'', Pauli's \cite{P} description of electron spin) is a previously
unrecognized essential ingredient in any attempt to construct a quantum hamiltonian
with eigenvalues corresponding to the imaginary parts of the nontrivial Riemann zeros.

Next, consider the ``completed'' zeta function $\Lam(s):=\Gi(s)\zeta(s)$, where 
$\Gi(s):=\pi^{-s/2}\G(s/2)$ and $\G(z)$ is the Euler gamma function.
The completed zeta function obeys Riemann's functional equation $\Lam(s)=\Lam(1-s)$,
and is real on the critical line; the zeros of $\Lam(s)$ coincide with the nontrivial zeros
of $\zeta(s)$.  It follows that the number of zeros of $\zeta(s)$ on the critical line
with imaginary part between zero and $E>0$ is given by
\beq
N(E) = {\textstyle\frac1\pi}\Im\log\Lam(\half+\eps+iE) + 1,
\label{NE}
\eeq
where $\eps$ is a positive infinitesimal \cite{fn}.
We can write $N(E)$ as the sum of a smooth contribution 
and an oscillating contribution \cite{B}:
\bea
N(E) &=& \Nbar(E)+\Nosc(E),
\label{NNN} \\
\Nbar(E) &=& {\textstyle\frac1\pi}\Im\log\Gi(\half+iE) + 1
\nonumber \\
&=& {\textstyle \frac{E}{2\pi}\log\!\left(\frac{E}{2\pi}\right)
-\frac{E}{2\pi}+\frac78 + O(E^{-1})},
\label{Nbar} \\
\Nosc(E) &=& {\textstyle\frac1\pi}\Im\log\zeta(\half+\eps+iE). 
\label{Nosc}
\eea
Using the Euler product formula $\zeta(s)=\prod_p(1-p^{-s})^{-1}$, where $p$ is a prime,
we get the formal expression
\bea
\Nosc(E) &=& -\frac{1}{\pi}\Im{\sum_p}\log(1-p^{-(1/2+iE)})
\nonumber \\
&=& +\frac{1}{\pi}\Im\sum_p\sum_{r=1}^\infty{\frac{p^{-r(1/2+iE/2)}}r}
\nonumber \\
&=& -\frac{1}{\pi}\sum_p\sum_{r=1}^\infty \frac{\sin(rE\log p)}{ r\,p^{r/2}}.
\label{Noscp}
\eea
This expression is formal because the Euler product does not converge on the critical line.
Its value is in its similarity to the corresponding expression for the number of energy 
eigenvalues less than $E$ of a hamiltonian for a classically chaotic system whose
classical periodic orbits are all isolated and unstable.  For a system without
the two-valued quantum degree of freedom required by class $C$,
the smooth contribution is given by the Weyl formula (see e.g.~\cite{H})
\beq
\Nbar(E)=\int \frac{d^f\!x\,d^f\!p}{(2\pi\hbar)^f}\;\Theta(0<h(x,p)<E),
\label{NbarQM}
\eeq
where $\Theta(S)=1$ if $S$ is true and $0$ if $S$ is false,
$f$ is the number of classical degrees of freedom, 
and $h(x,p)$ is the classical hamiltonian \cite{fn1}.
The oscillating contribution is given by a formal sum over primitive periodic orbits 
(labelled by $po$) and their repetitions (labelled by $r$),
\beq
\Nosc(E) = +\frac1{\pi\hbar}\sum_{po}\sum_{r=1}^\infty
\frac{\sin(rS_{po}/\hbar-r\mu_{po})}{r\,|\det(M_{po}^r-I)|^{1/2}},
\label{NoscQM}
\eeq
where the primitive orbit has action $S_{po}(E)$, Maslov phase $\mu_{po}(E)$, and stability matrix $M_{po}(E)$.

If we hypothesize a dynamical system in which the primitive periodic orbits are labelled 
by prime numbers \cite{B}, then \eq{NoscQM} bears a strong resemblance to \eq{Noscp}.
However, there are two well known problems with getting \eq{NoscQM} to reproduce \eq{Noscp} precisely \cite{B}.  
First, $|\det(M_{po}^r-I)|$ generically does not have the form of a simple
exponential like $p^r$.  Second, no value of $\mu_{po}$ in \eq{NoscQM}
will result in the overall minus sign on the right-hand side of \eq{Noscp}.

The generalization of \eq{NoscQM} to class $C$ has been considered by
Gnutzmann et al \cite{GSvOZ}.  As a prototypical class-$C$ system, 
they studied a Fermi sea of electrons (with the Fermi surface at $E=0$)
in a hard-wall billiard in a strong magnetic field (to break time-reversal invariance).
There are then both electron and hole excitations, and $\sigma_3$ is defined to be
$+1$ for electrons and $-1$ for holes.  Part of the billiard boundary is superconducting,
and this leads to Andreev reflection: when hitting the superconducting boundary, 
an electron turns into a hole (and vice versa) and ``retroflects'', initially retracing the incoming path.
There is an extra phase factor of $-i$ for each Andreev reflection, in addition to the Maslov phase.  
In general, the action of a primitive periodic orbit takes the form \cite{GSvOZ}
\beq
S_{po}(E)=S_{po}^{(e)}(E)+S_{po}^{(h)}(E),
\label{SpE}
\eeq
where $S_{po}^{(e)}(E)$ $[S_{po}^{(h)}(E)]$ is the action of those segments of the orbit 
where the excitation is an electron [hole].  For a given segment,
\beq
S_{\rm seg}^{(h)}(E) = -S_{\rm seg}^{(e)}(-E).
\label{Sseg}
\eeq
Gnutzmann et al show that the dominant periodic orbits are {\it self-dual\/}.  
A self-dual orbit includes an odd number
$N_A$ of Andreev reflections, and is traced twice, with each segment traced once as
an electron and once as a hole.  For a self-dual orbit, we therefore have
\beq
S_{po}(E) = S_{po}^{(e)}(E)-S_{po}^{(e)}(-E) \simeq E\tau_{po},
\label{Etaup}
\eeq
where $\tau_{po}=2\,\partial S_{po}^{(e)}/\partial E$ is the period of the complete twice-traced orbit.  
The Maslov phases of the two tracings cancel, 
but the factor of $-i$ for each Andreev reflection results in
an extra overall factor of $(-i)^{2N_A r}=(-1)^r$, where $r$ is the number of repetitions
of the complete orbit.  Finally, there are two factors of the inverse square-root
of the stability determinant, one for each single tracing.  The final result is therefore \cite{GSvOZ,fn2}
\beq
\Nosc(E) = \frac1{\pi\hbar}\sum_{po}\sum_{r=1}^\infty
\frac{(-1)^r\sin(r E \tau_{po}/\hbar)}{r\,|\det(M_{po}^r-I)|}.
\label{Nosc4}
\eeq

\Eq{Nosc4} bears a much stronger resemblance to \eq{Noscp} for the Riemann zeros than
does \eq{NoscQM}.  The dominant orbit actions are linear in $E$,
and the primitive orbits contribute with the correct sign.  

We can improve the agreement if we hypothesize that the underlying 
dynamical system has primitive periodic
orbits that are labelled by both a prime $p$ and another integer $k=0,1,\ldots$ 
(rather than by a prime $p$ alone),
and that, for a primitive orbit so labelled, $\tau_{po} = 2^k \log p$ and
$|\det(M_{po}^r-I)|=\exp(r\tau_{po}/2)$ \cite{so}.  
With this ansatz, we have (setting $\hbar=1$)
\bea
\Nosc(E) &=& \frac{1}{\pi}\sum_{po}\sum_{r=1}^\infty
\frac{(-1)^r\sin(r E \tau_{po})}{r\,|\det(M_{po}^r-I)|}
\nonumber \\
&=& \frac{1}{\pi}\sum_p\sum_{k=0}^\infty\sum_{r=1}^\infty
\frac{(-1)^r}{r}\frac{\sin(2^k r E \log p)}{\exp(2^{k} r \log p/2)}. \qquad
\label{Nosc5}
\eea
We now use the mathematical identity \cite{id}
\beq
\sum_{k=0}^\infty\sum_{r=1}^\infty \frac{(-1)^r}{r} f(2^k r) = -\sum_{r=1}^\infty \frac{1}{r} f(r).
\label{ff}
\eeq
Thus \eq{Nosc5} becomes
\beq
\Nosc(E) = -\frac{1}{\pi}\sum_p\sum_{r=1}^\infty
\frac{1}{r}\frac{\sin( r E \log p)}{\exp( r \log p/2)},
\label{Nosc6}
\eeq
which matches \eq{Noscp} precisely.  Thus, while the even repetitions contribute
with the wrong sign in \eq{Nosc4}, these contributions can in principle be balanced 
(in a class-$C$ system) by correct-sign contributions from other primitive orbits.  

Next we consider our results in comparison with some earlier work.

Connes \cite{Co} has suggested that the minus sign in \eq{Noscp} should be explained 
by having the Riemann zeros be {\it missing\/} eigenvalues in an otherwise continuous spectrum 
of an appropriate hamiltonian $H$.  
This would explain why all repetitions contribute with the same sign, but leaves open the
fundamental problem that matching Riemann zeros to missing eigenvalues does not allow
for a potential proof of the Riemann hypothesis by demonstrating that $\zeta(\half+iE)\propto\det(E-H)$.
Instead, Connes shows that the Riemann hypothesis is equivalent to a certain trace formula for
a hamiltonian with the desired continuous spectrum.  In the present work, we have provided an alternative explanation for the
sign discrepancy that still allows for the original formulation of the Hilbert--P\'olya conjecture.

Berry and Keating \cite{BK} have suggested that the quantum hamiltonian $H$ 
corresponding to the Riemann zeta function
should take the form of some quantization, on some compactified phase space for one  degree of freedom, 
of the classical hamiltonian $h(x,p)=xp$.  Here we note that this hamiltonian would be in 
class $D$.  To see this, consider the simplest hermitian quantization on an uncompactified
phase space, $H=\half(XP+PX)$, where $X$ and $P$ are the position and momentum operators.  
If we take matrix elements of this hamiltonian between basis states with real position-space wave functions,
we get a hamiltonian matrix of the form $H=A$, where $A$ is imaginary and antisymmetric.  
This characterizes hamiltonians in class $D$ \cite{AZ}.  
Class-$D$ systems have broken time-reversal invariance, and hence have eigenvalues
with a statistical distribution governed by GUE.  However, since a class-$D$ system does not have the
extra nonclassical two-valued degree of freedom, \eq{NoscQM} for $\Nosc(E)$ applies, 
and so the generic sign discrepancy with \eq{Noscp} is still present.

In conclusion, the combination of the Hilbert--P\'olya conjecture (that the imaginary parts of the nontrivial zeros of the Riemann zeta function are the eigenvalues of some quantum hamiltonian) with the Katz--Sarnak conjecture (that the Riemann zeta function is a member
of a family of $L$ functions related by a symplectic symmetry) implies that a hamiltonian
whose eigenvalues are the imaginary parts of the Riemann zeros should reside in 
class $C$ of the Altland--Zirnbauer classification scheme.  This implies that the hamiltonian should incorporate a nonclassical two-valued degree of freedom.  
Systems in class $C$ generically
have primitive periodic orbits that contribute to the density of the Riemann zeros with
the correct sign, further strengthening the argument that class $C$ is the right arena to
search for the elusive Riemann hamiltonian.

\acknowledgments
I thank Jeffrey Stopple for discussions and 
Sven Gnutzmann, Jon Keating, and Michael Berry for helpful correspondence.
This work was supported in part by the National Science Foundation under grant
PHY07-57035.


\begin{thebibliography}{99}
\bibitem{fn}
For small $z$, $\Lam(\half+z)=-c_1-c_2 z^2+O(z^4)$, where
$c_1$ and $c_2$ are positive real constants.  We choose the branch of the logarithm
such that $\Im\log\Lam(\half+\eps+iE)=-\pi$ for small positive $E$.   
As $E$ increases, this function jumps by $+\pi$ whenever $\Lam(\half+iE)$ changes sign.
\bibitem{fn1}
For a class-$C$ system, $h(x,p)$ is replaced in the Weyl formula by $a(x,p)+|\vec s(x,p)|$,
where $a(x,p)$ and $s_i(x,p)$ are the classical formulations of the operators $A$ and $S_i$.
\bibitem{so}
As noted earlier, $|\det(M_{po}^r-I)|$ does not generically have this simple exponential dependence on $r$.  This issue does not appear to be ameliorated solely by having $H$ 
be in class $C$, and so instead should emerge as a detailed property of the correct 
class-$C$ system.
\bibitem{id}
To verify \eq{ff}, set $n=2^k r$, and consider the terms on the left-hand side with a fixed value of $n$.
The possible values of $r$ are then $n,n/2,\ldots,n/2^m$, where $2^m$ is the largest power of 2
that divides $n$. Only the last of these possible values of $r$ is odd.  
Summing $(-1)^r/r$ over this set thus yields 
$(1/n)(1+2+\ldots+2^{m-1}-2^m)=-1/n$.  The remaining sum over $n$ is now the same as the sum 
over $r$ on the right-hand side of \eq{ff}.
\end{thebibliography}
\end{document}